\documentclass{PoS}

\usepackage{amsmath}
\usepackage{graphicx}
\newcommand{\Slash}[1]{{\ooalign{\hfil/\hfil\crcr$#1$}}}

\newcommand{\eqn}[1]{(\ref{#1})}

\def\simge{\mathrel{%
       \rlap{\raise 0.511ex \hbox{$>$}}{\lower 0.511ex \hbox{$\sim$}}}}
\def\simle{\mathrel{
       \rlap{\raise 0.511ex \hbox{$<$}}{\lower 0.511ex \hbox{$\sim$}}}}

\title{Equation of state in (2+1)-flavor QCD with gradient flow}

\ShortTitle{EOS in (2+1)-flavor QCD with gradient flow}

\author{\speaker{Kazuyuki Kanaya}
      \\
      CiRfSE, University of Tsukuba, Japan\\
      E-mail: \email{kanaya@ccs.tsukuba.ac.jp}
      }

\author{Shinji Ejiri\\
        Dept. of Physics, Niigata University, Japan\\
        E-mail: \email{ejiri@muse.sc.niigata-u.ac.jp}
        }

\author{Ryo Iwami\\
        Grad. School of Science and Technology, Niigata University, Japan\\
        E-mail: \email{iwami@muse.sc.niigata-u.ac.jp}
        }

\author{Masakiyo Kitazawa\\
        Dept. of Physics, Osaka University, Japan; J-PARC Branch, KEK Theory Center, KEK, Japan\\
        E-mail: \email{kitazawa@phys.sci.osaka-u.ac.jp}
        }

\author{Hiroshi Suzuki\\
        Dept. of Physics, Kyushu University, Japan\\
        E-mail: \email{hsuzuki@phys.kyushu-u.ac.jp}
        }

\author{Yusuke Taniguchi\\
        CCS, University of Tsukuba, Japan\\
        E-mail: \email{tanigchi@het.ph.tsukuba.ac.jp}
        }

\author{Takashi Umeda\\
        Grad. School of Education, Hiroshima University, Japan\\
        E-mail: \email{tumeda@hiroshima-u.ac.jp}
        }

\author{Naoki Wakabayashi\\
        Grad. School of Science and Technology, Niigata University, Japan\\
        E-mail: \email{wakabayashi@muse.sc.niigata-u.ac.jp}
        }

\abstract{
The energy-momentum tensor and equation of state are studied in finite-temperature $(2+1)$-flavor QCD with improved Wilson quarks using the method proposed by Makino and Suzuki based on the gradient flow. We find that the results of the gradient flow are consistent with the previous results using the $T$-integration method at $T \simle 280\,\mathrm{MeV}$ ($N_t\simge10$), while a disagreement is found at $T\simge350\,\mathrm{MeV}$ ($N_t \simle 8$) presumably due to the small-$N_t$ lattice artifact. 
We also report on the results on the renormalized chiral condensate and its disconnected susceptibility using the method of Hieda and Suzuki. The results show a clear signal of the expected chiral restoration crossover 
even with Wilson-type quarks which violate the chiral symmetry explicitly. }

\FullConference{34th annual International Symposium on Lattice Field Theory\\
		24-30 July 2016\\
		University of Southampton, UK}

\begin{document}

\section{Introduction}
\label{sec:intro}

Gradient flow is a kind of diffusion with a fictitious time $t$ and the flowed fields can be viewed as smeared fields over a range of about $\sqrt{8t}$ in four dimensions \cite{Narayanan:2006rf,Luscher:2009eq,Luscher:2010iy,reviewLattice}. 
Because operators constructed with the flowed fields are shown to be free from UV divergences and short-distance singularities, the gradient flow provides us with a new non-perturbative renormalization scheme with a physical scale of $\mu=1/\sqrt{8t}$.
This opened us a large variety of possibilities to drastically simplify the evaluation of physical observables on the lattice. 
In Ref.~\cite{flowqcd}, the energy-momentum tensor (EMT) and equation of state (EOS) in quenched QCD has been studied 
using the method of~\cite{Suzuki:2013gza} based on the gradient flow.
In this paper, we report on our study of finite-temperature $(2+1)$-flavor QCD with the gradient flow \cite{ourEOSpaper}. 
We compute EMT/EOS using the method of~\cite{Makino:2014taa} and the chiral condensate and its susceptibility using the method of~\cite{Hieda:2016lly}.


The gradient flow we adopt is the simplest one.
The gauge field is flowed as \cite{Luscher:2010iy}
\begin{equation}
   \partial_tB_\mu(t,x)=D_\nu G_{\nu\mu}(t,x),\quad
   B_\mu(t=0,x)=A_\mu(x),
\label{eq:(1.1)}
\end{equation}
where 
$
G_{\mu\nu}(t,x)\equiv
\partial_\mu B_\nu(t,x)-\partial_\nu B_\mu(t,x)
+
[B_\mu(t,x),B_\nu(t,x)],
$ and
\begin{equation}
D_\nu G_{\nu\mu}(t,x)\equiv
\partial_\nu G_{\nu\mu}(t,x)+[B_\nu(t,x),G_{\nu\mu}(t,x)].
\end{equation}
The quark fields are flowed by \cite{Luscher:2013cpa}
\begin{eqnarray}
&&
\partial_t\chi_f(t,x)=\Delta\chi_f(t,x),
\quad
\partial_t\Bar{\chi}_f(t,x)
   =\Bar{\chi}_f(t,x)\overleftarrow{\Delta},
\label{eq:(1.15)}
\end{eqnarray}
where $ \chi_f(t=0,x)=\psi_f(x)$, $\Bar{\chi}_f(t=0,x)=\Bar{\psi}_f(x)$, and $f=u$, $d$ and $s$, with 
\begin{eqnarray}
&&
\Delta\chi_f(t,x)\equiv D_\mu D_\mu\chi_f(t,x),
\quad
D_\mu\chi_f(t,x)\equiv\left[\partial_\mu+B_\mu(t,x)\right]\chi_f(t,x),
\notag\\&&
\Bar{\chi}_f(t,x)\overleftarrow{\Delta}
   \equiv\Bar{\chi}_f(t,x)\overleftarrow{D}_\mu\overleftarrow{D}_\mu,
\quad
   \Bar{\chi}_f(t,x)\overleftarrow{D}_\mu
   \equiv\Bar{\chi}_f(t,x)\left[\overleftarrow{\partial}_\mu-B_\mu(t,x)\right].
\end{eqnarray}
This simple flow dependent only on gauge fields requires wave function renormalization of the quark fields, but besides it the finiteness of the flowed operators is preserved \cite{Luscher:2013cpa}.


We use $(2+1)$-flavor QCD configurations generated for Refs.~\cite{Ishikawa:2007nn,Umeda:2012er}
with non-perturbatively $O(a)$-improved Wilson quarks and Iwasaki glue~\cite{Iwasaki:2011np}. 
Our gauge coupling $\beta=2.05$ corresponds to~$a=0.0701(29)\,\mathrm{fm}$, 
and our hopping parameters correspond to $m_\pi/m_\rho\simeq0.63$ and $m_{\eta_{ss}}/m_\phi\simeq0.74$. The bare PCAC
quark masses are $a\,m_{ud}=0.02105(17)$ and $a\,m_s=0.03524(26)$.
Using the fixed-scale approach~\cite{Levkova:2006gn,Umeda:2008bd}, we study lattices with $N_t=16$, 14, $\cdots$ 4 corresponding to $T=1/(aN_t)\simeq 174$, 199, $\cdots$ 697 MeV, 
where the pseudo critical temperature is $T_\mathrm{pc}\sim 190$ MeV \cite{Umeda:2012er}.
Spatial box size is $32^3$ for finite temperature and $28^3$ for zero temperature.
To avoid unphysical effects due to overlapped smearing, $\sqrt{8t}$ should be smaller than the half of the lattice extents: 
\begin{equation}
t \le t_{1/2} \equiv \frac{1}{8}\left[\min\left(\frac{N_t}{2},\frac{N_s}{2}\right)\right]^2.
\label{eq:t-half}
\end{equation}
See Ref.~\cite{ourEOSpaper} for further details of the simulation parameters and algorithms, including the temperature and number of configurations at each $N_{t}$.

\section{Energy-momentum tensor and equation of state}
\label{sec:EMT}

The correctly normalized EMT is given by \cite{Suzuki:2013gza,Makino:2014taa}
\begin{align}
   T_{\mu\nu}(x)
   &=\lim_{t\to0}\biggl\{c_1(t)\left[
   \Tilde{\mathcal{O}}_{1\mu\nu}(t,x)
   -\frac{1}{4}\Tilde{\mathcal{O}}_{2\mu\nu}(t,x)
   \right]
   +c_2(t)\left[
   \Tilde{\mathcal{O}}_{2\mu\nu}(t,x)
   -\left\langle\Tilde{\mathcal{O}}_{2\mu\nu}(t,x)\right\rangle_{\! 0}
   \right]
\notag\\
   &
   +c_3(t)\sum_{f=u,d,s}
   \left[
   \Tilde{\mathcal{O}}_{3\mu\nu}^f(t,x)
   -2\Tilde{\mathcal{O}}_{4\mu\nu}^f(t,x)
   -\left\langle
   \Tilde{\mathcal{O}}_{3\mu\nu}^f(t,x)
   -2\Tilde{\mathcal{O}}_{4\mu\nu}^f(t,x)
   \right\rangle_{\! 0}
   \right]
\label{eq:(2.8)}\\
   &
   +c_4(t)\sum_{f=u,d,s}
   \left[
   \Tilde{\mathcal{O}}_{4\mu\nu}^f(t,x)
   -\left\langle\Tilde{\mathcal{O}}_{4\mu\nu}^f(t,x)\right\rangle_{\! 0}
   \right]
   +\sum_{f=u,d,s}c_5^f(t)\left[
   \Tilde{\mathcal{O}}_{5\mu\nu}^f(t,x)
   -\left\langle\Tilde{\mathcal{O}}_{5\mu\nu}^f(t,x)\right\rangle_{\! 0}
   \right]\biggr\},
\notag
\end{align}
with $\langle\cdots\rangle_{0}$ standing for the expectation value at zero-temperature 
and
\begin{align}
  &
   \Tilde{\mathcal{O}}_{1\mu\nu}\equiv
   G_{\mu\rho}^a\,G_{\nu\rho}^a,
   \qquad{}
   \Tilde{\mathcal{O}}_{2\mu\nu}\equiv
   \delta_{\mu\nu}\,G_{\rho\sigma}^a\,G_{\rho\sigma}^a,
   \qquad{}
   \Tilde{\mathcal{O}}_{3\mu\nu}^f\equiv
   \varphi_f(t)\,\Bar{\chi}_f
   \left(\gamma_\mu\overleftrightarrow{D}_\nu
   +\gamma_\nu\overleftrightarrow{D}_\mu\right)
   \chi_f,
\notag\\
   &
   \Tilde{\mathcal{O}}_{4\mu\nu}^f\equiv
   \varphi_f(t)\,\delta_{\mu\nu}\,
   \Bar{\chi}_f
   \overleftrightarrow{\Slash{D}}
   \chi_f,
   \qquad{}
   \Tilde{\mathcal{O}}_{5\mu\nu}^f\equiv
   \varphi_f(t)\,\delta_{\mu\nu}\,
   \Bar{\chi}_f\,
   \chi_f,
\label{eq:(2.13)}
\end{align}
where
$
   \overleftrightarrow{D}_\mu\equiv D_\mu-\overleftarrow{D}_\mu,
$
and the quark normalization factor~$\varphi_f(t)$ is given by~\cite{Makino:2014taa}
\begin{equation}
   \varphi_f(t)\equiv
  - \frac{6}
   {(4\pi)^2\,t^2
   \left\langle\Bar{\chi}_f(t,x)\overleftrightarrow{\Slash{D}}\chi_f(t,x)
   \right\rangle_{\! 0}}.
\label{eq:(2.15)}
\end{equation}
We then have $p/T^4  = \sum_i \langle T_{ii}\rangle /(3T^4)$, $\epsilon/T^4 = - \langle T_{00}\rangle/T^4$ for EOS.
The coefficients $c_i(t)$ 
are given in~\cite{Makino:2014taa}.
Though these coefficients are calculated by perturbation theory, they are used just to guide the $t\to0$ extrapolation. We thus consider that our evaluation is essentially non-perturbative. 
We also emphasize that the non-perturbative beta functions or Karsch coefficients, which require a big numerical task in conventional calculations of EOS in particular in full QCD, are not needed. 

The extrapolation $t\to0$ is required to remove contamination of unwanted operators.
In \eqn{eq:(2.8)}, the continuum extrapolation $a\to0$ is assumed to be done before.
In numerical studies, however, it is often favorable to take the continuum extrapolation at a later stage of analyses.
On finite lattices with $a\ne0$, we expect additional contamination of unwanted operators.
Since we adopt the non-perturbatively $O(a)$-improved Wilson quarks,
the lattice artifacts start with $O(a^2)$.
We expect
\begin{align}
   T_{\mu\nu}(t,x,a)
   &=T_{\mu\nu}(x) + t\,S_{\mu\nu}(x)
   +A_{\mu\nu}\frac{a^2}{t}+\sum_f B_{f\mu\nu}\,(am_f)^2+C_{\mu\nu}\,(aT)^2
   \notag\\
&\qquad{}
   +D_{\mu\nu}\left(a\Lambda_{\mathrm{QCD}}\right)^2
   +a^2S'_{\mu\nu}(x)+O(a^4,t^2),
\label{eq:a2overt}
\end{align}
where $T_{\mu\nu}(x)$ is the physical EMT, $S_{\mu\nu}$ and $S'_{\mu\nu}$ are contaminations of dimension-six operators with the same quantum number, and 
$A_{\mu\nu}$, $B_{f\mu\nu}$, $C_{\mu\nu}$, and~$D_{\mu\nu}$ are those from dimension-four operators. 
To exchange the order of limiting procedures $a\to0$ and $t\to0$, the singular terms like $a^2/t$ have to be removed. 
This is possible if we have a window in $t$ in which the linear terms of \eqn{eq:a2overt} dominate.

\begin{figure}[t]
\begin{center}
\includegraphics[width=7cm]{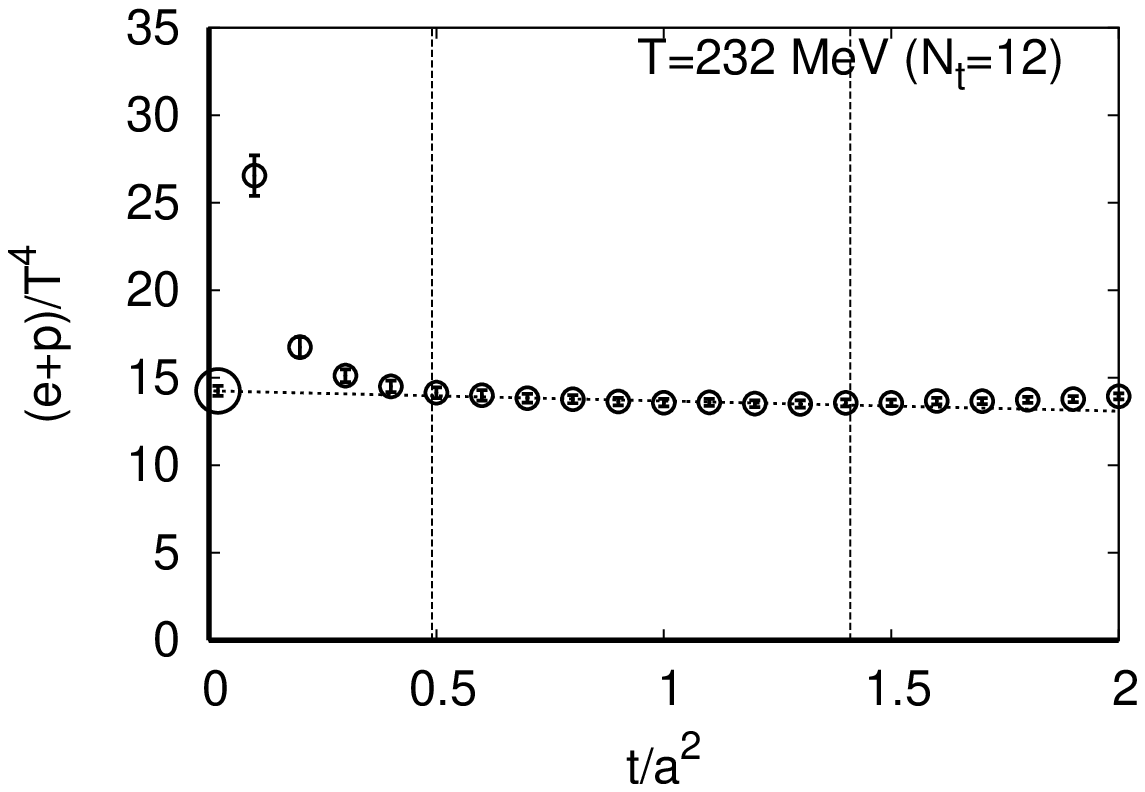}
\includegraphics[width=7cm]{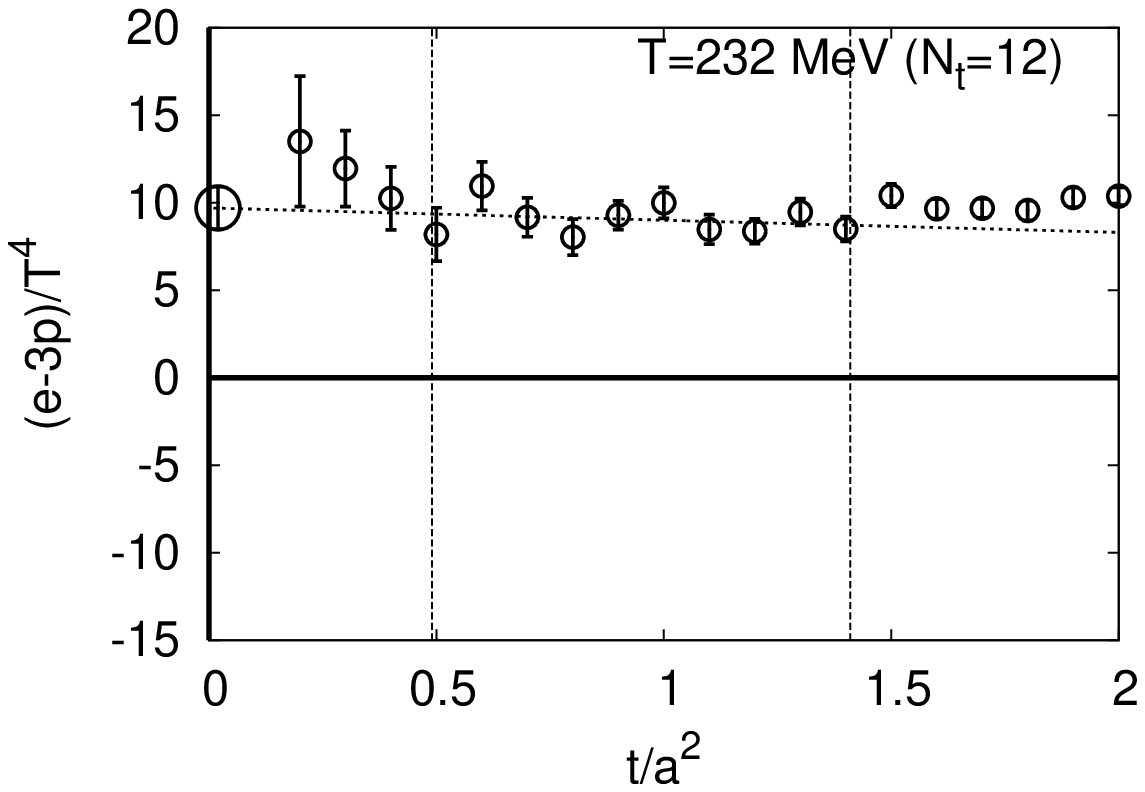}
\vspace{-5mm}
\caption{Entropy density (\textbf{left}) and trace anomaly (\textbf{right}) at $T\simeq232$ MeV as function of the flow time~$t/a^2$. Pair of dotted vertical lines indicates the window used for the linear fit.  Errors are statistical only.}
\label{eos1}
\end{center}
\end{figure}

\begin{figure}[t]
\begin{center}
\includegraphics[width=7cm]{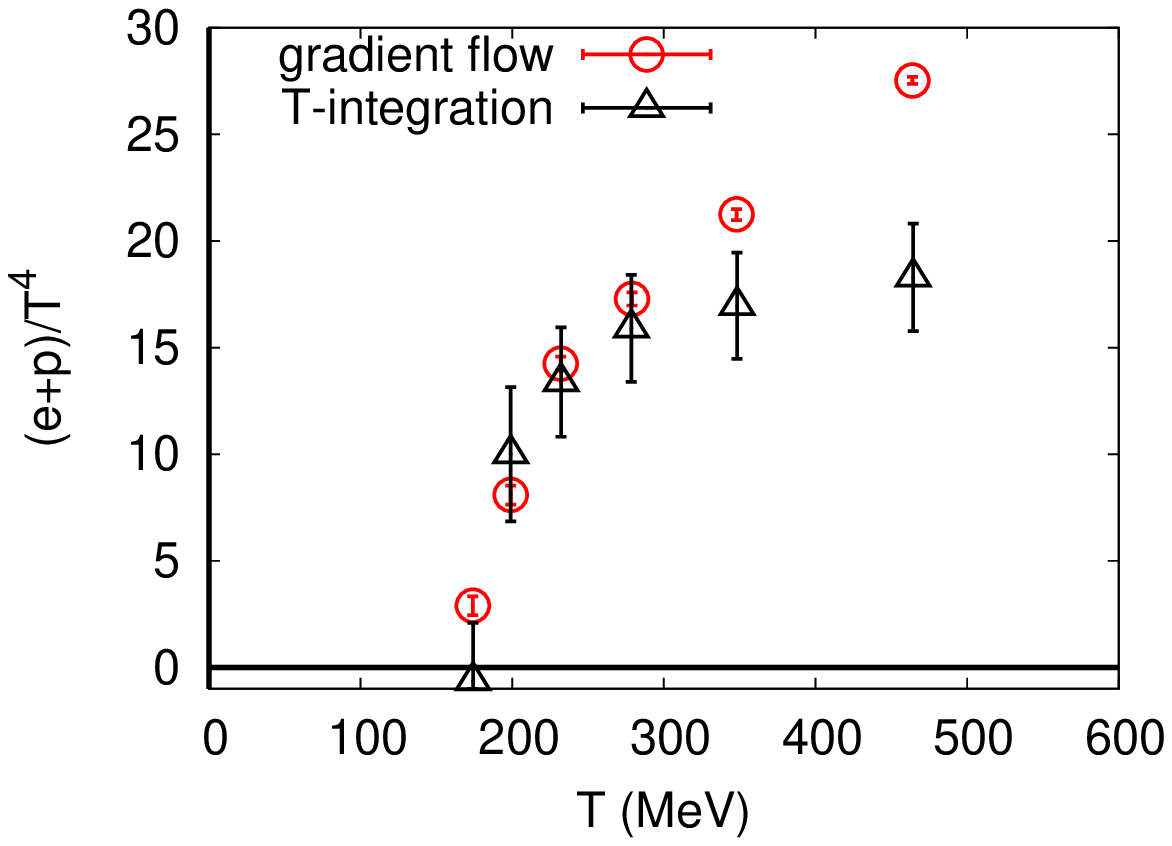}
\includegraphics[width=7cm]{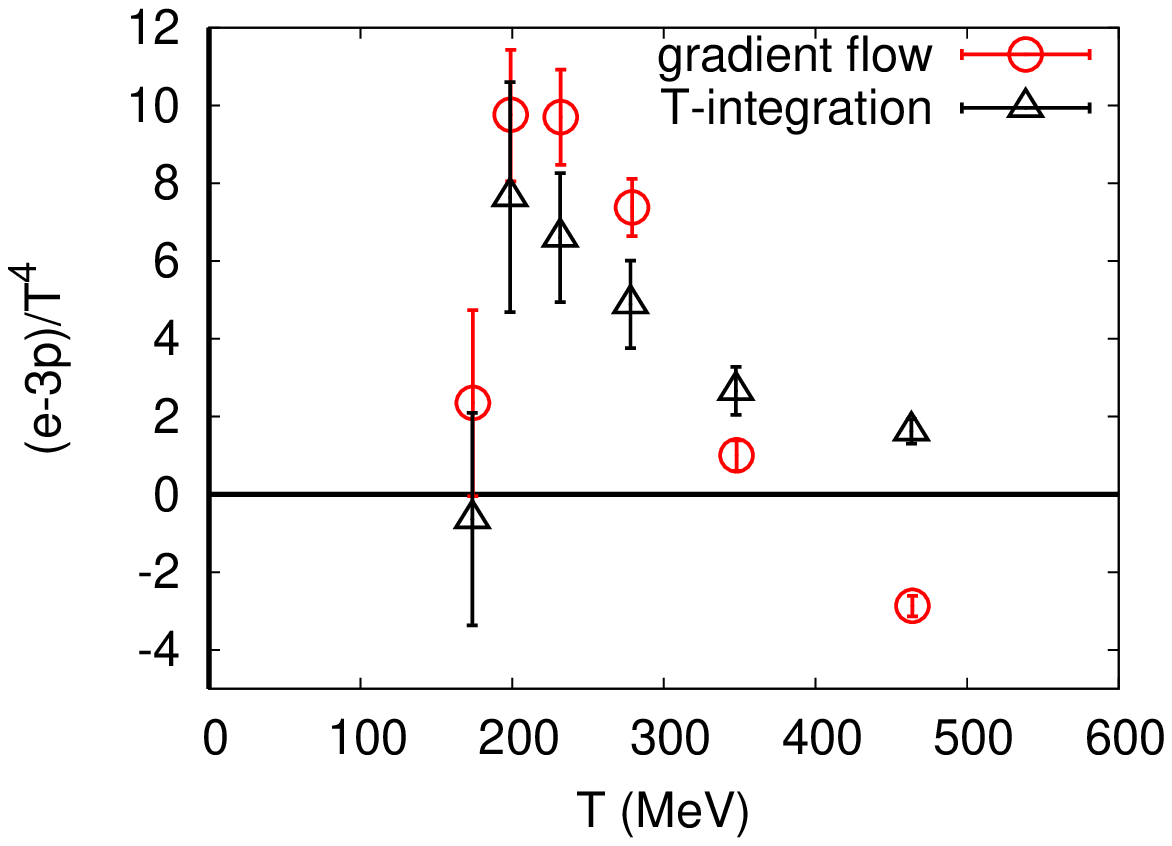}
\vspace{-3mm}
\caption{Final results for entropy density (\textbf{left}) and trace anomaly (\textbf{right}) as function of temperature. Red circles are our result using the gradient flow.  
Errors include the statistical error and the systematic error from the perturbative coefficients.
Black triangles are the previous result by the $T$-integration method~\cite{Umeda:2012er}.}
\label{eos2}
\end{center}
\end{figure}

In Fig.~\ref{eos1}, we show the results of the entropy density $(\epsilon+p)/T^4$ and the trace anomaly $(\epsilon-3p)/T^4$  at $T\simeq232$ MeV as functions of $t/a^2$.
We find windows below $t_{1/2}$ in which the data is well linear.
Results at other $T$'s are similar, except for the case of $T\simeq697$ MeV ($N_t=4$) for which $t_{1/2}=0.5$ and no linear windows are found below $t_{1/2}$.
We perform linear extrapolations $t\to0$ using the data in the linear window at $T\simle497$ MeV to get red circles in Fig.~\ref{eos2}.
The black triangles are the results of our previous study of EOS with the $T$-integration method~\cite{Umeda:2012er} using the same configurations.
We find that the result of the gradient flow method is well consistent with that of the conventional method at~$T\simle279\,\mathrm{MeV}$. 
On the other hand, the two results deviate at $T\simge348\,\mathrm{MeV}$ ($N_t\simle8$). 
This may be due to that the lattice artifact of $O\left((aT)^2\right)=O\left(1/N_t^2\right)$ in \eqn{eq:a2overt} is non-negligible for $N_t\simle8$.%
\footnote{
At $T\simeq464$ MeV ($N_t=6$), because $t_{1/2}=1.125$ for this lattice is quite small, the window for linear extrapolation is narrow. 
We also note that $(\epsilon-3p)/T^4$ has small curvature above this $t_{1/2}$~\cite{ourEOSpaper}.
The negative $(\epsilon-3p)/T^4$ at this $T$ shown in Fig.~\ref{eos2} may be suggesting that, besides the small-$N_t$ lattice artifacts, the true linear region is hidden above $t_{1/2}$.} 
%

\section{Chiral condensate and its disconnected susceptibility}
\label{sec:chiralcond}

We then study the chiral condensate.
Although the Wilson-type quarks violates the chiral symmetry explicitly, 
the gradient flow method~\cite{Hieda:2016lly} enables us to directly evaluate 
the chiral condensate and its susceptibility on the lattice. 
We define the scalar density as the chiral rotation of the pseudo-scalar density whose normalization is uniquely fixed by the PCAC relation.
The flowed chiral condensate at small $t$ and $a$ is expected to be \cite{Hieda:2016lly}
\begin{align}
   \left\{\Bar{\psi}_f\psi_f\right\}(t,x,a)
   &\equiv 
   c_s(t)\,
   \frac{\Bar{m}_f\!\left(1/\sqrt{8t}\right)}{m_f}
   \left[\varphi_f(t)\,\Bar{\chi}_f(t,x)\,\chi_f(t,x)\right]    
\notag\\
   &=
 \left\{\Bar{\psi}_f\psi_f\right\}_{\overline{\mathrm{MS}}}(x)
   +\frac{m_f}{t}N
   +t\,S'(x)
   +A\frac{a^2}{t}
   +\sum_fB_f\,(am_f)^2
+C\,(aT)^2
\notag\\
   &\qquad{}      +D\left(a\Lambda_{\mathrm{QCD}}\right)^2
+a^2S(x)+O(a^4,t^2) ,
\label{eqn:bpsipsi-a}
\\
c_s(t) &\equiv 
   1+\frac{\Bar{g}\!\left(1/\sqrt{8t}\right)^2}{(4\pi)^2}
   \left[4(\gamma-2\ln2)+8+\frac{4}{3}\ln(432)\right]
\end{align}
where $f=u$, $d$ and $s$.
The coefficient $c_s(t)$ transforms the gradient flow renormalization scheme to ${\overline{\mathrm{MS}}}$ scheme 
such that $\left\{\Bar{\psi}_f\psi_f\right\}_{\overline{\mathrm{MS}}}(x)$ is the renormalized chiral condensate in ${\overline{\mathrm{MS}}}$ scheme at $\mu=2$ GeV. 
$N$~is a contamination of dimensionless operators,
$S'$ and $S$ are from dimension-five operators,  and 
$A$, $B_f$, $C$, and $D$ are from dimension-three operators.

Note that, in addition to the $a^2/t$ term, $m_f/t$ term appears off the chiral limit. 
Existence of the $m_f/t$ term originates from the fact that the singlet part of the scalar density possesses the quantum number identical to the vacuum. In fact, we obtain such term by the lowest order perturbation theory at small $t$. 
The $m_f/t$ term is an obstacle to take the $t\to0$ limit even in the continuum limit. 
To overcome the problem, Ref.~\cite{Hieda:2016lly} suggests to subtract the $T=0$ expectation value (VEV subtraction). 
At $a\ne0$, we have in any case the $a^2/t$ term. 
We test the chiral condensate both with and without the VEV subtraction.

\begin{figure}[t]
\begin{center}
\includegraphics[width=7cm]{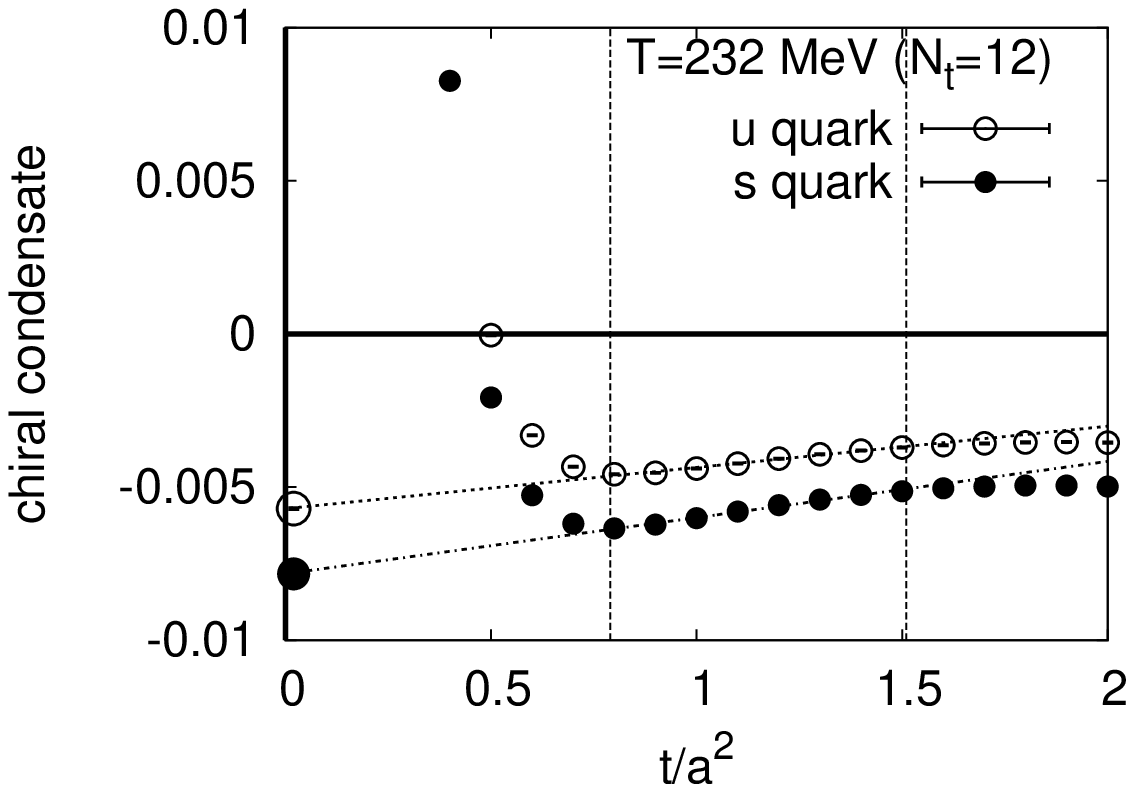}
\includegraphics[width=7cm]{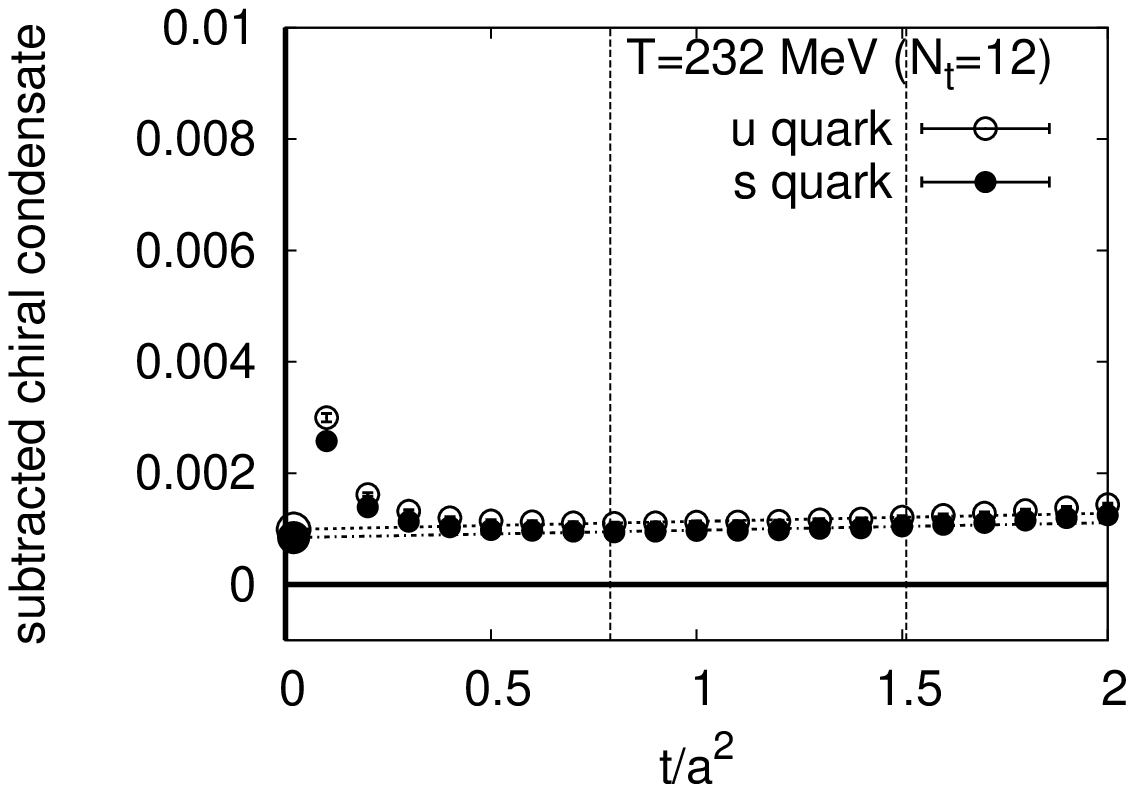}
\vspace{-5mm}
\caption{Unsubtracted (\textbf{left}) and VEV subtracted (\textbf{right}) chiral condensate at $T\simeq232$ MeV as function of the flow time $t/a^2$. The vertical axis is in lattice unit.  
Open and filled circles are for $u$ (or $d$) and $s$ quarks, respectively. 
Dotted vertical lines show the window for the linear fit.
Errors are statistical only.}
\label{chi1}
\end{center}
\end{figure}

\begin{figure}[t]
\begin{center}
\includegraphics[width=7cm]{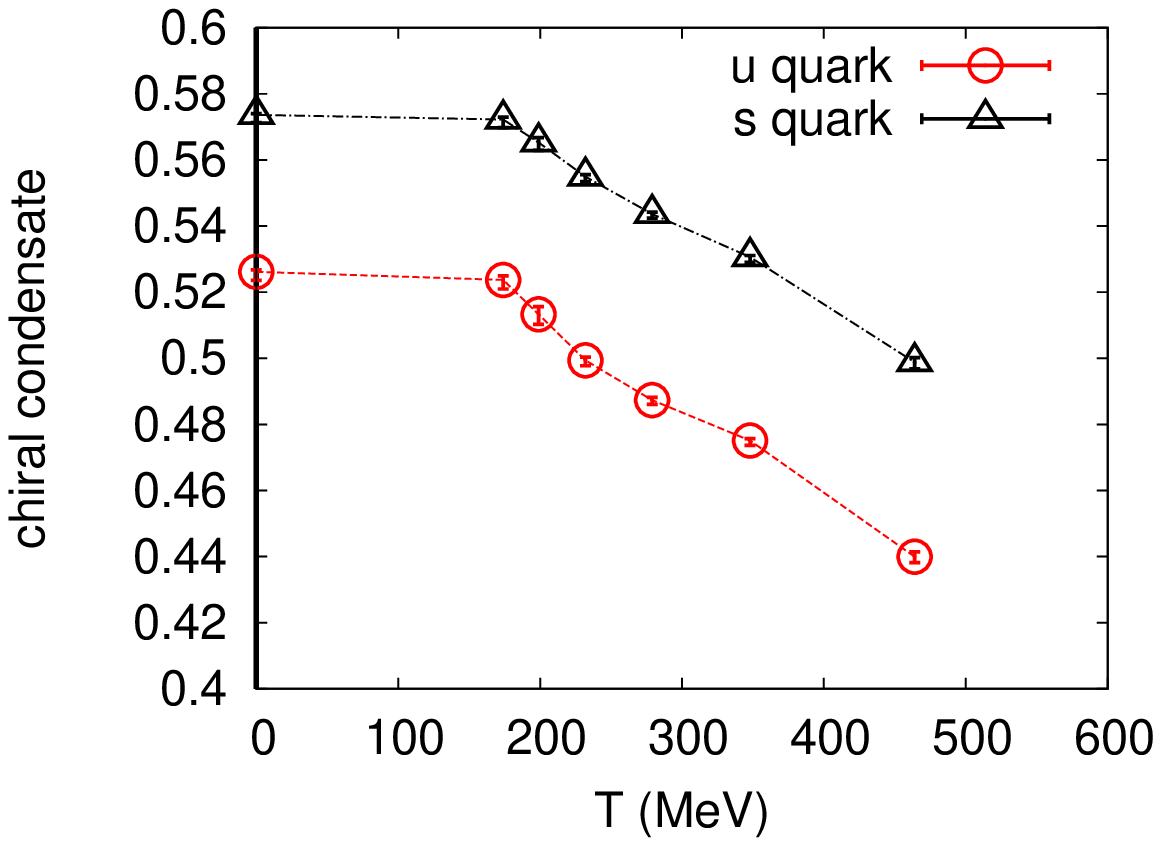}
\includegraphics[width=7cm]{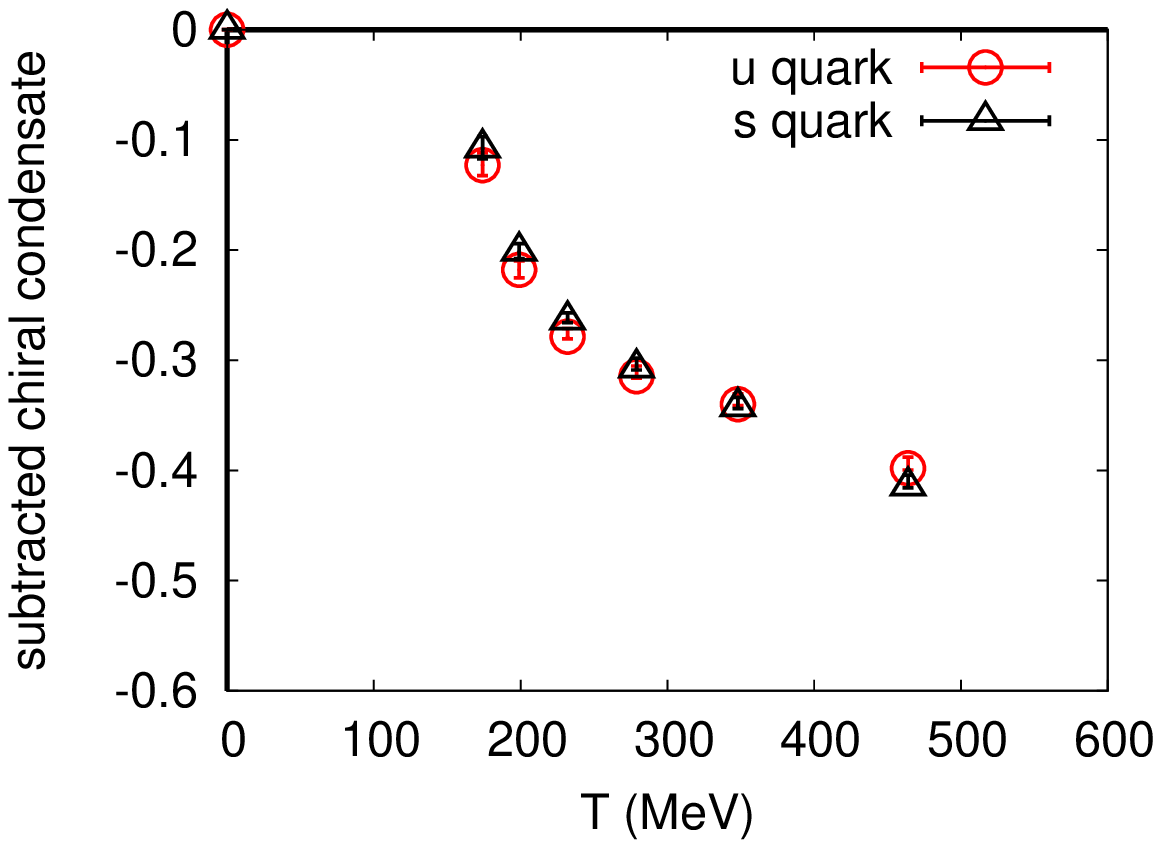}
\vspace{-2mm}
\caption{Renormalized chiral condensate $-\left\langle\{\Bar{\psi}\psi\}(x)\right\rangle_{\overline{\mathrm{MS}}}(\mu\!=\!2{\mathrm{GeV}})$
 in $\overline{\mathrm{MS}}$ scheme as function of temperature,
obtained from the $t\to0$ limit of the unsubtracted (\textbf{left}) and VEV subtracted (\textbf{right}) chiral condensates. 
The vertical axis is in unit of GeV. Red circles are $u$ (or $d$) quark condensate and
black triangles are that for $s$ quark. 
Errors include the statistical error and the systematic error from the perturbative coefficients.}
\label{chi2}
\end{center}
\end{figure}

We show in Fig.~\ref{chi1} the chiral condensate at $T\sim232$ MeV as function of $t/a^2$.
Similar to the case of EMT, we find linear windows below $t_{1/2}$ in which the $a^2/t$ and $m_f/t$ terms as well as higher orders in $t$ seem to be well suppressed.
Results at other $T$'s are also similar except  
at $T\simeq697$ MeV ($N_t=4$) where no window was found below $t_{1/2}$. 
Carrying out linear $t\to0$ extrapolations at $T\simle464$ MeV adopting the same fit range as EMT, we obtain Fig.~\ref{chi2} for the renormalized chiral condensate without and with the VEV subtraction.
The condensates seem to start change at $T\sim190$ MeV,
in accordance with a previous estimate of $T_{\mathrm{pc}}$~\cite{Umeda:2012er}. 
The unsubtracted chiral condensate shows a tendency to
decrease as we decrease the valence quark mass. The behavior is consistent with 
our expectation.
With the VEV subtraction, the condensate becomes almost independent of the valence quark mass, i.e.\ the mass-dependent part is almost $T$-independent.

\begin{figure}[t]
\begin{center}
\includegraphics[width=7cm]{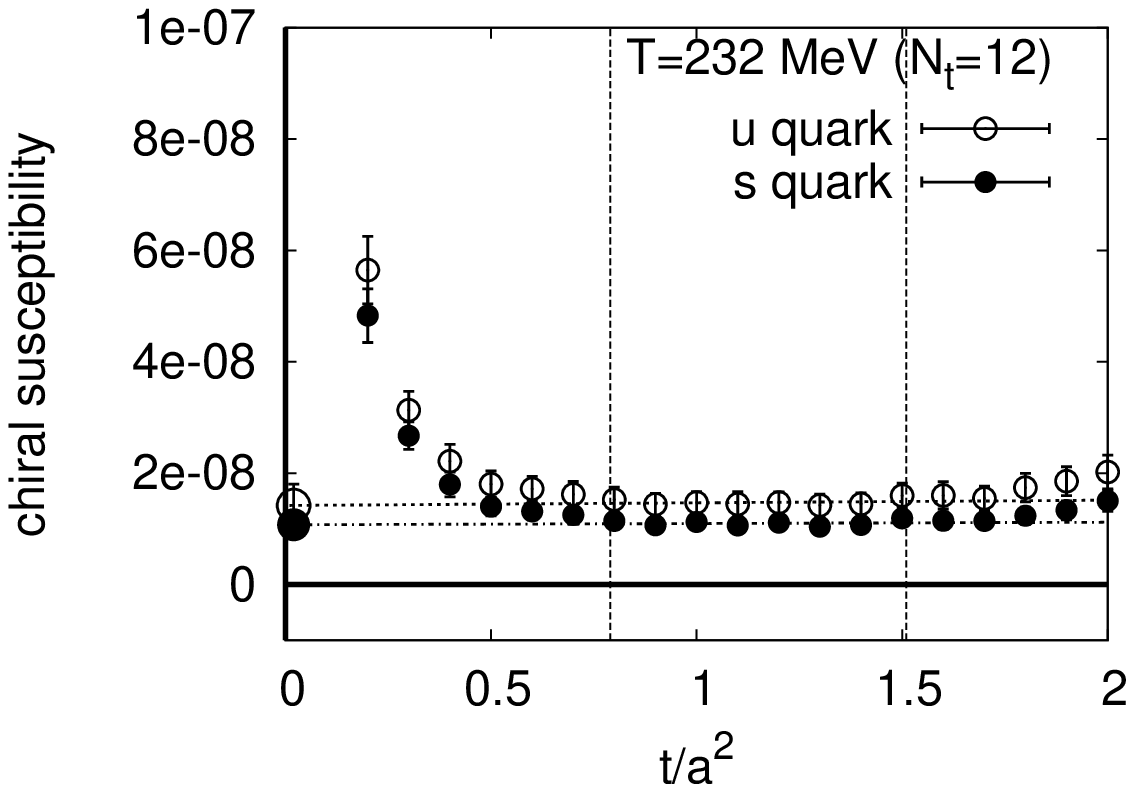}
\includegraphics[width=7.5cm]{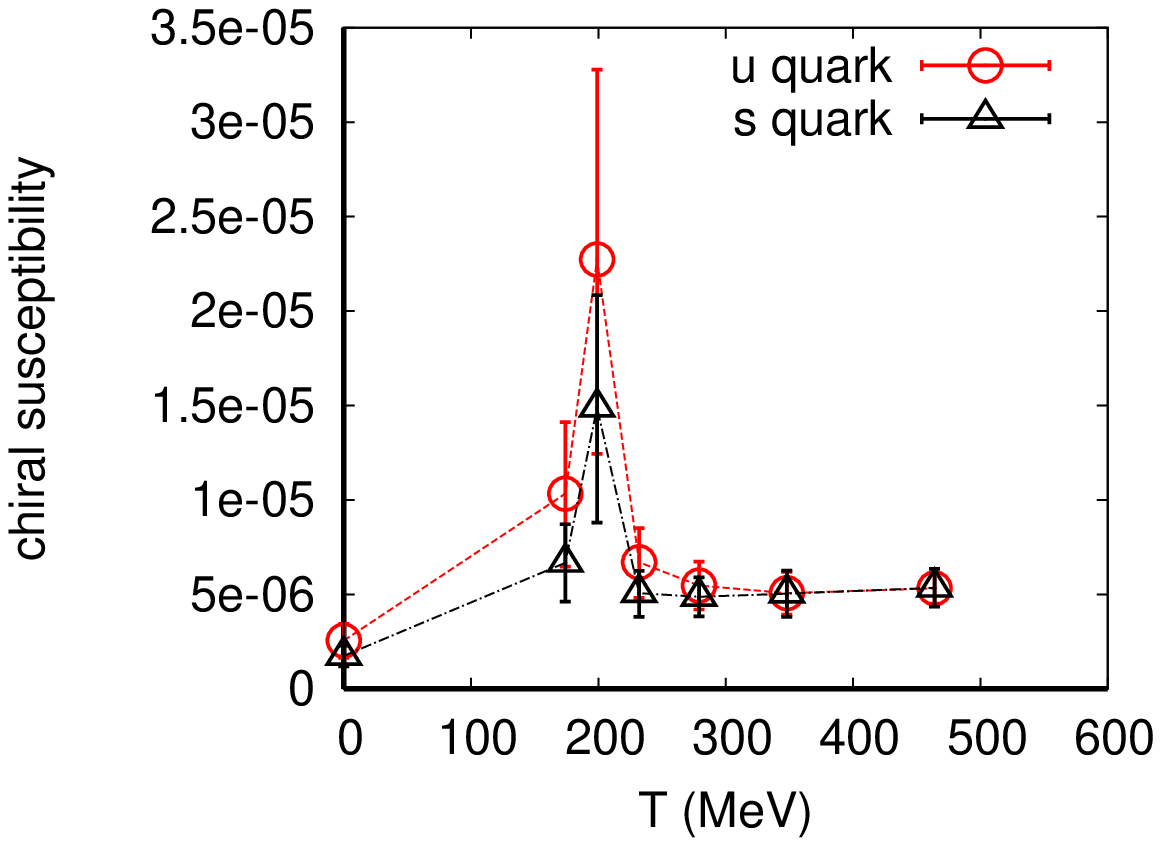}
\vspace{-3mm}
\caption{\textbf{Left:} The same as Fig.~3 but for disconnected chiral susceptibility. In this quantity, the VEV subtraction has no effects.
\textbf{Right:} Renormalized disconnected chiral susceptibility in $\overline{\mathrm{MS}}$ scheme at $\mu=2$ GeV as a function of temperature. The vertical axis is in unit of $\mathrm{GeV}^6$. Red circles are for $u$ (or $d$) quark and black triangles are for $s$ quark. Errors include the statistical error and the systematic error from the perturbative coefficients.}
\label{chi3}
\end{center}
\end{figure}

We now calculate the disconnected chiral susceptibility,
\begin{equation}
   \chi_{\Bar{f}f}^{\mathrm{disc.}}
   \equiv \left\langle \left[ \frac{1}{N_\Gamma} \sum_x \{\Bar{\psi}_f\psi_f\}(x) \right]^2 \right\rangle_{\!\mathrm{disconnected}}
   -\; \left[ \left\langle\frac{1}{N_\Gamma} \sum_x \{\Bar{\psi}_f\psi_f\}(x)\right\rangle \right]^2.
\end{equation}
This quantity is not a physical susceptibility but may be used as a guide to detect the chiral restoration transition.
In the left panel of Fig.~\ref{chi3}, the disconnected chiral susceptibility at $T\simeq232$ MeV is shown as function of the flow time.
We find a good linear window. 
Results at other $T$'s are agin similar to the case of EMT.
Linear $t\to0$ extrapolations at $T\simle464$ MeV lead to the right panel of Fig.~\ref{chi3}.
We find a clear peak of the susceptibility at around $T\sim190$ MeV.
The peak hight shows a tendency to increase as we decrease the valence quark mass.

\section{Conclusions and discussions}
\label{sec:conclusion}

We calculated EOS and chiral condensate in $(2+1)$-flavor QCD with improved Wilson quarks at a single but fine lattice spacing applying the gradient flow
method of Refs.~\cite{Suzuki:2013gza,Makino:2014taa,Hieda:2016lly}.
We found that the results of EOS by the gradient flow method are consistent
with those of the $T$-integration method at $T\simle280\,\mathrm{MeV}$ ($N_t \simge 10$). Although the continuum extrapolation is not done yet, the good agreement between the completely different methods suggests that our lattices are already close to the continuum limit.
On the other hand, deviation found at $T\simge350\,\mathrm{MeV}$ suggests that the lattice artifacts of $O\left((aT)^2\right)=O\left(1/N_t^2\right)$ from the discretization of thermal modes are not negligible at $N_t \simle 8$. 
%
%
The chiral condensate and its disconnected susceptibility show a clear signal of the chiral restoration crossover, even with the Wilson-type quarks.
These results demonstrates that the gradient flow is quite powerful in extracting physical observables from lattice simulations.

\vspace{3mm}
This work is in part supported by JSPS KAKENHI Grant
No.\ 25800148, No.\ 26287040, No.\ 26400244, No.\ 26400251, No.\ 15K05041,
and No.\ 16H03982,
by the Large Scale Simulation Program of High Energy Accelerator
Research Organization (KEK) No.\ 14/15-23, 15/16-T06, 15/16-T-07, and 15/16-25, 
and by Interdisciplinary Computational Science Program in CCS, University of
Tsukuba.
This work is in part based on Lattice QCD common code Bridge++ \cite{bridge}.


\begin{thebibliography}{99}

\bibitem{Narayanan:2006rf} 
  R.~Narayanan and H.~Neuberger,
  JHEP {\bf 0603}, 064 (2006).

\bibitem{Luscher:2009eq} 
  M.~L\"uscher,
  Commun.\ Math.\ Phys.\  {\bf 293}, 899 (2010).

\bibitem{Luscher:2010iy} 
  M.~L\"uscher,
  JHEP {\bf 1008}, 071 (2010);
  Erratum: [JHEP {\bf 1403}, 092 (2014)].

\bibitem{reviewLattice}
M.~L\"uscher, 
 PoS LAT {\bf 2013}, 016 (2014). 

\bibitem{flowqcd}
   M.~Kitazawa, T.~Iritani, M.~Asakawa, T.~Hatsuda, H.~Suzuki,
   arXiv:1610.07810 [hep-lat].


\bibitem{Suzuki:2013gza} 
  H.~Suzuki,
  PTEP {\bf 2013}, 083B03 (2013)
  Erratum: [PTEP {\bf 2015}, 079201 (2015)].


\bibitem{ourEOSpaper}
Y.~ Taniguchi, S.~Ejiri, R.~Iwami, K.~Kanaya, M.~Kitazawa, H.~Suzuki,
T.~Umeda, and N.~Wakabayashi,
  arXiv:1609.01417 [hep-lat].


\bibitem{Makino:2014taa} 
  H.~Makino and H.~Suzuki,
  PTEP {\bf 2014}, 063B02 (2014)
  Erratum: [PTEP {\bf 2015}, 079202 (2015)].

\bibitem{Hieda:2016lly} 
  K.~Hieda and H.~Suzuki,
  arXiv:1606.04193 [hep-lat].



\bibitem{Luscher:2013cpa} 
  M.~L\"uscher,
  JHEP {\bf 1304}, 123 (2013).


\bibitem{Ishikawa:2007nn} 
  T.~Ishikawa {\it et al.} [CP-PACS and JLQCD Collaborations],
  Phys.\ Rev.\ D {\bf 78}, 011502(R) (2008).

\bibitem{Umeda:2012er}
  T.~Umeda {\it et al.} [WHOT-QCD Collaboration],
  Phys.\ Rev.\ D {\bf 85}, 094508 (2012).

\bibitem{Iwasaki:2011np}
  Y.~Iwasaki,
  arXiv:1111.7054 [hep-lat].

\bibitem{Levkova:2006gn}
  L.~Levkova, T.~Manke and R.~Mawhinney,
  Phys.\ Rev.\ D {\bf 73}, 074504 (2006).

\bibitem{Umeda:2008bd}
  T.~Umeda, S.~Ejiri, S.~Aoki, T.~Hatsuda, K.~Kanaya, Y.~Maezawa and H.~Ohno,
  Phys.\ Rev.\ D {\bf 79}, 051501 (2009).




\bibitem{bridge}
{\tt http://bridge.kek.jp/Lattice-code/index\_e.html}


\end{thebibliography}
\end{document}